\documentclass[%
 reprint,
superscriptaddress,
twocolumn,
 amsmath,amssymb,
 aps,
pra,
]{revtex4-2}

\usepackage[T1]{fontenc}
\usepackage{graphicx}
\usepackage{dcolumn}
\usepackage{bm}
\usepackage{hyperref}
\usepackage[dvipsnames]{xcolor}
\usepackage{chemformula}
\usepackage{soul}
\usepackage{nameref}

\begin{document}

\title{Superconducting meander-line surface coil for NMR spectroscopy of nanoscale thin films}

\author{L. Beaudoin}
\affiliation{Département de physique and Institut quantique, Université de Sherbrooke, Sherbrooke, Quebec J1K2R1, Canada}

\author{A. Verrier}
\affiliation{Département de physique and Institut quantique, Université de Sherbrooke, Sherbrooke, Quebec J1K2R1, Canada}

\author{Y. A. Bioud}
\affiliation{Département de génie électrique et de génie informatique, Université de Sherbrooke, Sherbrooke, Québec, J1K 2R1, Canada}

\author{M. Massicotte}
\affiliation{Département de physique and Institut quantique, Université de Sherbrooke, Sherbrooke, Quebec J1K2R1, Canada}
\affiliation{Département de génie électrique et de génie informatique, Université de Sherbrooke, Sherbrooke, Québec, J1K 2R1, Canada}

\author{B. Reulet}
\affiliation{Département de physique and Institut quantique, Université de Sherbrooke, Sherbrooke, Quebec J1K2R1, Canada}

\author{J. A. Quilliam}
\affiliation{Département de physique and Institut quantique, Université de Sherbrooke, Sherbrooke, Quebec J1K2R1, Canada}

\date{\today}

\begin{abstract}
Nuclear magnetic resonance (NMR) spectroscopy is a powerful technique to study local magnetism in a variety of materials. However, the inherently low sensitivity of conventional inductively detected solid state NMR typically requires a large number of spins, reducing its applicability to two-dimensional (2D) materials and nanoscale thin films. To overcome this experimental challenge, we introduce a novel probe based on a superconducting meander-line surface coil that significantly enhances the NMR sensitivity for thin samples. Using a NbN meander with an optimized geometry, we demonstrate the sensitivity of this technique by detecting the NMR signal of a 150-nm-thick boron film containing only $\sim 2\times10^{16}$  $^{11}$B nuclear spins. Spin-echo measurements and theoretical modeling offer insight into the parameters limiting the coil's performance.  This work lays the foundation for developing highly sensitive NMR probes, potentially unlocking new opportunities for studying atomically thin materials. 
\end{abstract}

\maketitle

The study of thin films and 2D materials is an important branch of materials science, driven in part by the miniaturization of electronic devices and the exploration of novel condensed matter phenomena. However, their reduced dimensions pose a significant challenge for conventional measurement techniques that are designed to study macroscopic samples. NMR spectroscopy is one such technique, with widespread applications as a local probe of magnetic structures, susceptibility and fluctuations in solid state systems.~\cite{berthier_2017, walstedt_2008, slichter_1996, lacroix_2011} Yet, the small magnetic moment of nuclear spins severely limits the sensitivity of this technique when studying small-volume samples. To overcome this issue, many adaptations have been proposed~\cite{Lee2015}. For example, magnetic resonance force microscopy~\cite{lee_2007, won_2013}, nitrogen-vacancy centers in diamond~\cite{balasubramanian_2008,maze_2008} and dynamical nuclear polarization~\cite{abragam_1978,walder_2019} have been used to enhance NMR sensitivity. While effective, these techniques require far more complex instrumentation than a standard NMR spectrometer and are furthermore incompatible with many experimental conditions.

A more straightforward strategy that requires minimal or no modifications to existing instrumentation is to adapt the geometry of a standard NMR coil to match the sample dimensions. This approach enhances the signal from small samples by increasing the filling factor $\eta$ of the coil \cite{hoult_1976}, which is defined as the ratio of the RF magnetic energy confined within the sample volume to the total magnetic energy stored in the coil. Various coils have been realized to probe small samples, including microcoils~\cite{Badilita2010,Ehrmann2007,Peck1995,Lacey1999,lepucki_microcoils_2021}, micro-strip lines~\cite{Jasinski2012} and surface coils~\cite{Ackerman1990,massin_2002,buess_1991,liu_2017,Ali2017}. Lepucki \textit{et al.} have conducted a comprehensive review of various 3d coil geometries and their associated limits of detection~\cite{Lepucki2021}.

In the case of 2d thin-film samples, however, an ideal configuration is provided by meander-line surface coils. Their geometry generates an RF magnetic field that decays exponentially with the distance $x$ from the meander surface as $\exp(-2\pi x/a)$, where $a$ is the meander period~\cite{buess_1991}. Therefore, by reducing $a$ and placing the sample in proximity to the meander where the RF field is strongly confined, a much higher filling factor can be achieved compared to solenoidal microcoils or circular surface coils. Notably, Liu \textit{et al.} have demonstrated NMR signal detection from a 400-$\mu$m-thick \ch{InP} sample using a meander-line surface coil with a period $a\sim1$ mm\cite{liu_2017}. This approach has even been extended to probe $\sim 2\times 10^{16}$ $^{77}$Se spins in a 210-monolayer \ch{FeSe} film~\cite{lou_2023}. In principle, a better signal-to-noise for very thin samples can be achieved by further reducing the meander period, thereby confining the RF field even closer to the coil surface. However, for meanders made of normal metals, decreasing $a$ leads to higher resistance and a lower quality factor, ultimately degrading the SNR \cite{hoult_1976} and significantly limiting their applicability to the study of thin films.

\begin{figure*}
    \centering
    \includegraphics[width=\textwidth]{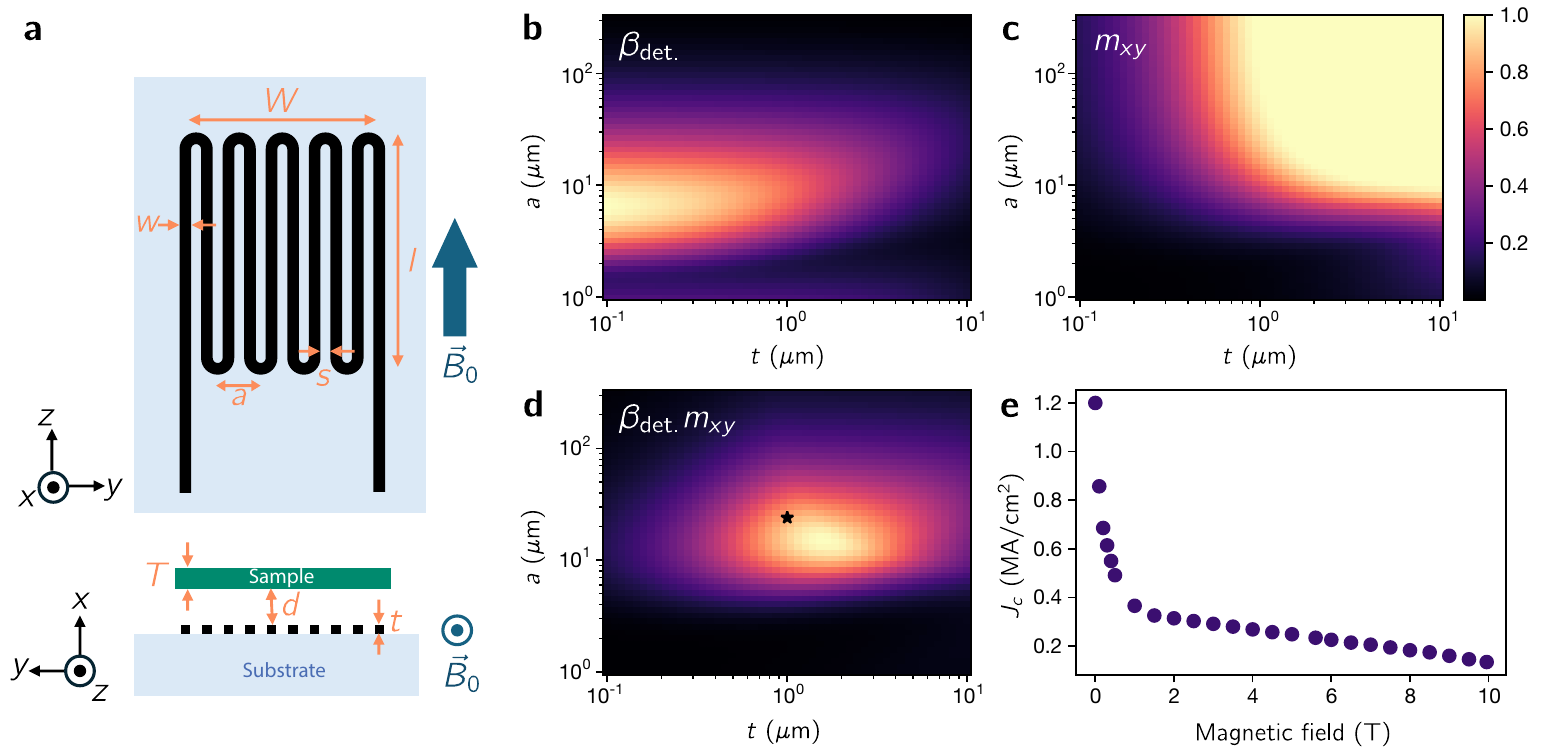} 
    \caption{{\bf a} A schematic (not to scale) of the meander-line inductance used in this work, showing the device parameters, the sample position and the orientation of the magnetic field $\vec{B}_0$.  
    {\bf b} The  detection efficiency $\beta_\mathrm{det.}$, {\bf c} the transverse magnetization  $m_{xy}$ (or excitation efficiency) and {\bf d} the product $\beta_\mathrm{det.} m_{xy}$, calculated as a function of wire thickness $t$ and meander period $a$. The star indicates the geometry used in our measurements. {\bf e} Critical current density $j_c$ measured on a NbN stripe with width $w = 3.5~\mu$m and thickness $t=1~\mu$m. }
    
    \label{fig:Optimization}
\end{figure*}

Here we propose the use of a lithographic superconducting meander-line surface coil for NMR excitation and detection in thin-film samples. The use of a superconductor enables a sub-micron meander period $a$ while preserving a sufficiently high quality factor. However, superconducting films also present challenges due to their sensitivity to magnetic fields. To mitigate this issue, we selected NbN, a type-II superconductor with relatively high critical temperature ($\sim 14$~K), field ($\sim 40$~T) and current density ($\sim 1$ MA/cm$^2$) \cite{Gavaler1981}. Taking these constraints into account, we optimized the geometry of the coil to maximize its sensitivity for nanoscale thin films. Finally, we demonstrate NMR measurements on a 150-nm-thick $^{11}$B thin film and discuss a strategy to extend this technique to atomically thin samples.

\section{Model and optimization}\label{sec:theory} 

The sensitivity of NMR coils is typically characterized by the SNR of the detected signal \cite{hoult_1976}. In designing our coil, we assume that its quality factor is sufficiently high such that the dominant source of noise in our setup is controlled by the preamplifier. This condition is readily achieved at low temperatures using a superconducting coil. Thus, our goal is to optimize the dimensions of our meander-line coil to maximize SNR by enhancing the NMR signal $\xi$. Fig.~\ref{fig:Optimization}a displays the general design of our probe, highlighting the key parameters to optimize: the thickness of the NbN film $t$, the width $w$ and length $l$ of the wires, the total width $W$ of the meander, and its period $a$. A sample of thickness $T$ is placed at distance $d$ from the coil.

In our experiment, the external magnetic field $\vec{B}_0$ is applied along the meander lines ($z$-axis). A current $\mathcal{I}$ flowing through the meander generates a magnetic field $\vec{B}_1$, which is predominantly oriented perpendicular to $\vec{B}_0$ with magnitude $B_{xy}(\vec{r})$ at position $\vec{r}$, generating a component $B_1 = B_{xy}/2$ that is static in the rotating reference frame. Following a RF pulse of duration $t_A$ at the resonant frequency $\omega_0$, the nuclear magnetic moments at position $\vec{r}$ are tipped by an angle $\theta(\vec{r}) = \langle I,m-1 | \hat{I}_- | I, m\rangle \gamma t_AB_1(\vec{r}) =  \sqrt{I(I+1)-m(m-1)} \gamma t_A B_1(\vec{r})$.\cite{fukushima_1981} Given that our test sample is a thin-film of boron, we consider the central line ($m=1/2$) of the spin-3/2 $^{11}$B nucleus which gives $\theta(\vec{r}) = 2\gamma t_A B_1(\vec{r})$.  According to the principle of reciprocity and neglecting spin relaxation, the voltage amplitude $\xi$ induced in the meander by the precession of the nuclear magnetic moments is given by~\cite{Ackerman1990} 
\begin{equation} \label{eq:1} \xi = \omega_0 \int_{V_S} M_0 \sin\theta(\vec{r}) b_1(\vec{r}) d^3\vec{r} \end{equation} 
where $M_0$ is the nuclear magnetization, $V_S$ is the sample volume and $b_1(\vec{r}) = B_1(\vec{r})/\mathcal{I}$ is the RF magnetic field per unit current in the meander-line inductor.

If $b_1$ is relatively homogeneous across the sample, $\xi$ can be approximated by
\begin{equation} \xi \simeq \omega_0 M_0V_S m_{xy} \beta_\mathrm{det.} \end{equation}
where 
\begin{equation} \beta_\mathrm{det.} = \frac{1}{V_S} \int_{V_S} b_1(\vec{r}) d^3\vec{r} \end{equation}
is the average current-normalized RF field in the sample volume and
\begin{equation} m_{xy} = \frac{M_{xy}}{M_0} = \frac{1}{V_S} \int_{V_S} \sin \theta(\vec{r}) d^3\vec{r} \end{equation}
represents the the average projection of the nuclear magnetization in the $xy$-plane. These quantities characterize the efficiency of the probe’s two main functions: excitation and detection. As such, they can serve as key figures of merit for analyzing and optimizing its performance.

To calculate $m_{xy}$ and $\beta_\mathrm{det.}$, we consider a number of experimental factors related to our test sample,  a boron film with a thickness of 150 nm. This material was chosen due to its relatively high nuclear gyromagnetic ratio (similar to many isotopes of interest for solid-state NMR), modest quadrupolar moment, ease of deposition and low conductivity (thereby avoiding a detrimental effect on the quality factor of the resonant circuit). Additionally, boron is otherwise absent from our NMR setup, preventing unwanted background signals. Based on prior studies, we expect a linewidth of approximately 50 kHz.~\cite{Turner2015}. We therefore limit our calculations to pulse lengths $t_A \leq 20$~$\mu$s, ensuring the entire spectrum can be irradiated with a single central frequency. Moreover, we assume that, for practical reasons, the sample is positioned at $d\sim 1 \mu$m away from the meander surface, such that $d \gg T$. We further constrain the parameter space of our optimization problem by considering that the wire width $w$ and spacing $s$ are equal. Finally, we use Biot-Savart law to calculate the RF field, assuming uniform current density in the superconducting strips. While this model ignores the effect of the London penetration depth and skin depth on the current distribution, we find that more sophisticated models~\cite{sethares_1977,sethares_1978} provide very similar results for a reasonable range of parameters. 

The factor $m_{xy}$ is related to the tip angle and is optimized either with the duration or the magnitude of the RF pulses. However, using overly long pulses to achieve a full 90$^\circ$ tip angle leads to a loss of bandwidth, extending the time required to obtain a full spectrum. On the other hand, the magnitude of the RF pulse that can be applied to our coil is limited by the critical current density $j_c$  of the superconductor. The measured $j_c$ of our NbN thin film, as a function of magnetic field, is shown in Fig.~\ref{fig:Optimization}e. We note that $j_c$ decreases appreciably with $B_0$ as a result of dissipation by vortices, but that the superconducting state is nonetheless preserved even for $B_0>10$ T. Fig.~\ref{fig:Optimization}c presents calculations for $m_{xy}$ as a function of the meander period $a$ and wire thickness $t$, assuming  $j_c \sim 0.2$~MA/cm$^2$. As expected, $m_{xy}$ is optimized for thicker and wider wires (large $a$ and $t$) where larger RF currents can be applied to achieve a full 90$^\circ$ tip angle. 

However, increasing the dimensions of the wires and the meander period reduces the magnitude of $b_1$ at the surface of coil, and thus its detection sensitivity. This trade-off is illustrated in Fig.~\ref{fig:Optimization}b which shows calculations of $\beta_\mathrm{det.}$ for the same parameters. We find that it is maximal for a period $a \simeq 6.5$ $\mu$m. This coincides with the optimal value $a = 2\pi d$ that can be analytically derived by modeling the meander with a sinusoidal 2D current distribution. In addition, calculations indicate that minimizing the wire thickness $t$ enhances $\beta_\mathrm{det.}$, as the current in a thicker meander is on average farther from the sample. 

The result of the trade-off between $\beta_\mathrm{det.}$ and $m_{xy}$ is given by their product, as shown in Fig.~\ref{fig:Optimization}d. From this we conclude that the NMR signal $\xi \propto \beta_\mathrm{det.} m_{xy}$ is maximal with a period of $a \sim 15$ $\mu$m and a thickness of $t \sim 2$ $\mu$m. Optimization of $\xi$ using Eq.\ref{eq:1} yields the same results, confirming that the inhomogeneity of $b_1$ is negligible. Finally, the total area $A = Wl$ of the meander is chosen such that its inductance $L$ is comparable to that of the standard NMR coil of roughly $L \sim 0.2~\mu$H, allowing us to cover a wide frequency range using a conventional NMR probe. Taking into account the contribution of the self-inductance and mutual inductance of the wires ~\cite{mohan_1999,greenhouse_1974,stojanovic_2006}, this results in an area of approximately 1.5 mm$^2$.

\begin{figure*}
    \centering
    \includegraphics[width=\textwidth]{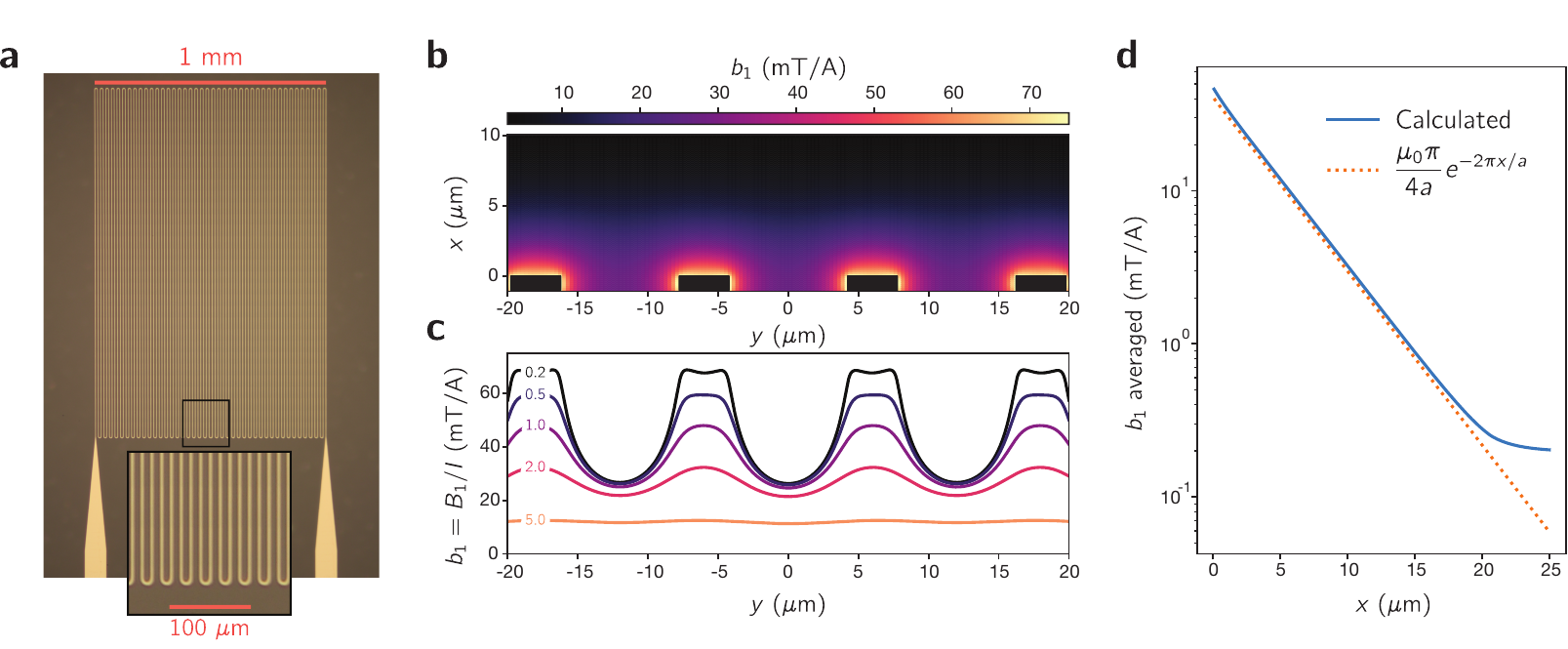}
    \caption{\textbf{a} Optical image of the the meander-line surface coil fabricated with NbN sputtered on sapphire. \textbf{b}, \textbf{c} RF field per unit of current $b_1$ in the vicinity of the meander calculated assuming a uniform current distribution. The values used for this calculation correspond to those of the meander shown in \textbf{a}. \textbf{d} Exponential decrease of $b_1$ as a function of distance from the sample $x$. The solid curve is the calculated average RF field as a function of $x$ and the dotted line is calculated using the equation  $ b_1 = \frac{\mu_0 }{8\lambda} e^{-x/\lambda}$ with $\lambda = a/2\pi$ and $a= 24$~$\mu$m. }
    \label{fig:OurDevice}
\end{figure*}

\section{Experiment}\label{sec:Exp_setup}

Guided by these optimization results, we fabricated the meander-line surface coil shown in Fig.~\ref{fig:OurDevice}a. Some compromises were made, in particular regarding the film thickness, to simplify the fabrication. A $t=1$ $\mu$m-thick layer of \ch{NbN} was deposited by reactive sputtering on a sapphire substrate. The patterning of the meander-lines was done by photolithography and reactive ion etching. The resulting meander consists of 84 parallel wires $w = 3.5$~$\mu$m wide and $l = 1.5$ mm long with a period $a = 24$~$\mu$m, for a total meander width $W$ of 1 mm and area of 1.5 mm$^2$. The test sample was fabricated by sputtering a 150 nm-thick layer of boron onto an undoped silicon substrate. To maximize the sensitivity of the probe, the sample must be positioned close to the coil at a distance $d \ll a$. This was achieved by using a $\sim 1$ $\mu$m-thick poly(methyl methacrylate) (PMMA) layer as adhesive between the sample and meander.

Fig.~\ref{fig:OurDevice}b shows calculations of the current-normalized magnetic field $b_{xy}$ generated by this coil. At a distance $x = d = 1$ $\mu$m from the meander, corresponding to the sample position, the average field is estimated to be $b_1 \simeq 35$ mT/A, with a homogeneity of $\pm9$ mT/A. As can be seen from Fig.~\ref{fig:OurDevice}c, the field becomes less homogeneous closer to the meander. As predicted the averaged RF field is strongly confined to the vicinity of the meander surface and decays exponentially as a function of the distance from the meander $x$ with a characteristic length $\lambda = a/2\pi$. Near $x \sim a$ the exponential decay gives way to power-law drop arising from the finite size of the meander.

The coil and sample were placed inside a standard NMR probe, connected in series and parallel with two variable capacitors, each ranging from 1 and 20 pF. Given the estimated inductance of our coil ($L \sim 0.2\mu$H ), this RF circuit can operate over a frequency range of 80 to 360 MHz with a good impedance matching and quality factor ($Q>$100). All measurements were performed at 4.2 K, well below the critical temperature $T_c \sim 12$ K of NbN. RF pulses were generated by a TecMag \emph{Redstone} console  and delivered to the NMR probe through the coupled port of a directional coupler. The NMR signal was amplified at the output of the coupler by a RF preamplifier (Miteq AU-1114) and analyzed by the console. 

As previously noted, the RF power that can be applied to our sample is inherently limited by the critical current of the superconducting coil. This constraint simplifies our NMR setup as it eliminates the need for a high-power RF amplifier and duplexer, which are typically required in conventional systems. In any case, the larger coupling between the surface coil and the sample (large $b=B/I$ ratio) means that we can nearly obtain optimal tip angles using exclusively the output of the NMR console with 0 dB of amplification. The pulse lengths that provided the maximum signal were found to be 13 and 7 $\mu$s at 5.6 and 10 T, respectively. 

\section{Results}\label{sec:Results}

\begin{figure*}
    \centering
    \includegraphics[width=\textwidth]{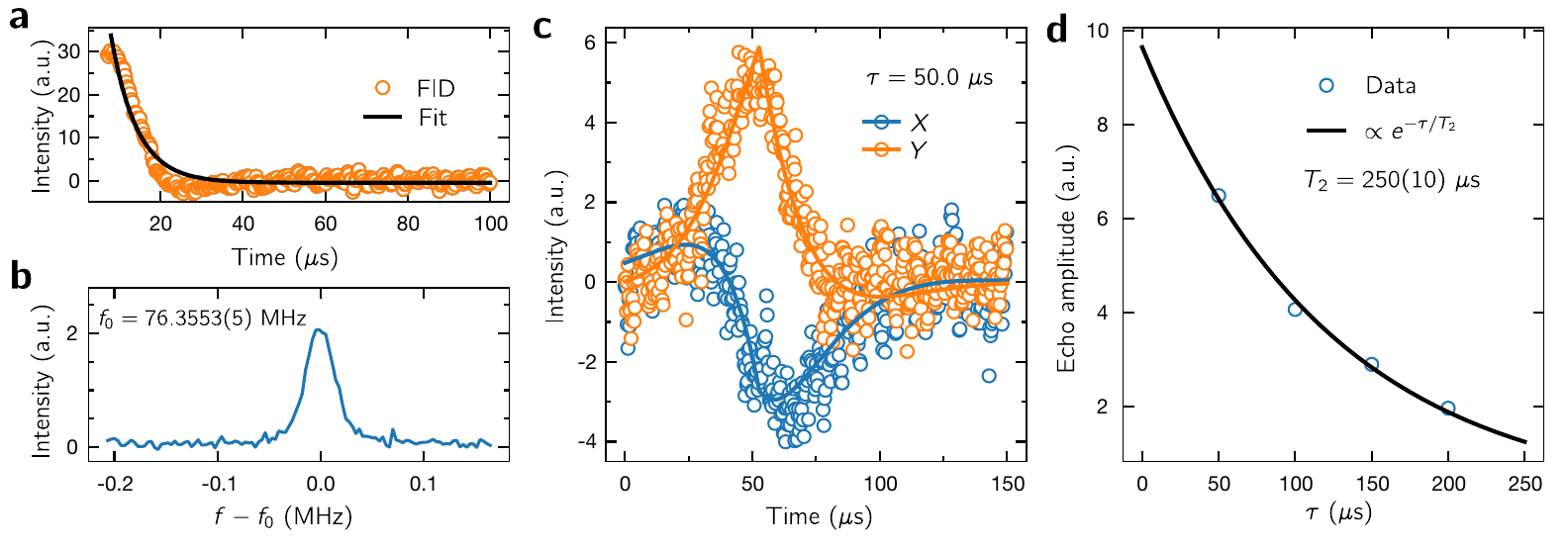}
    \caption{$^{11}$B NMR signal measured from a 150-nm-thick boron thin film with an area of 1.5 mm$^2$ at $T_s =4.2$~K in a static magnetic field of 5.6~T.   \textbf{a} Magnitude of the free induction decay (FID). This measurement was obtained after $4.4\times 10^5$ scans, with $N = 1024$ data points per scan and a dwell time of 300 ns (receiver bandwidth $\Delta f = 3.33$ MHz). An exponential fit provides a crude estimate of $T_2^\ast$. \textbf{b} Fourier transform following a single pulse. \textbf{c} Spin-echo signal and corresponding fit. \textbf{d} Integrated spin-echo signal as a function of $\tau$, with an exponential fit used to extract the value of $T_2$. }
    \label{fig:Results}
\end{figure*}

Fig.~\ref{fig:Results} shows the successful detection of a $^{11}$B NMR signal using our meander-line inductance set-up. First we present a basic free-induction decay (FID) following a single excitation pulse. The modulus of the signal is presented in Fig.~\ref{fig:Results}a along with an exponential fit giving an approximate value of $T_2^\ast \simeq 10.7$ $\mu$s. To accurately extract $S_\mathrm{FID}$, the exponential decay is shifted in time so that $t=0$ corresponds to the center of the RF excitation pulse of length $t_A$~\cite{fukushima_1981}. The Fourier transform of the FID, shown in Fig.~\ref{fig:Results}b, exhibits a linewidth (FWHM) of 36 kHz, in agreement with values previously reported in the literature.\cite{Turner2015} The ratio $f_0/B_0$ is measured to be $13.635\pm 0.003$ MHz/T at 5.6 T and $13.629\pm 0.001$ MHz/T at 9.95 T. These values are very close to the reported value of the gyromagnetic ratio of $^{11}$B, $^{11}\gamma / 2\pi = 13.6629846(5)$ MHz/T.\cite{stone_2005} The small discrepancy is a result of not measuring an accurate reference signal and instead relying on the accuracy of the magnet power supply. However, the difference is minimal, leaving no doubt that the detected signal originates from $^{11}$B NMR in the sample. 

To make this coil truly useful for solid-state NMR applications, it is important to achieve large tip angles  and implement spin-echo measurements~\cite{hahn_1950}. Fig.~\ref{fig:Results}c displays the spin-echo signal obtained following a standard $t_A$-$\tau$-$2t_A$ sequence, where $t_A$ is the length of the first pulse. The optimal pulse length $t_A$, ranging between 7 to 13 $\mu$s depending on the applied field, is determined experimentally. This measurement is repeated for different waiting times ($\tau$), with the integrated signal strength $S_\mathrm{Echo}(\tau)$ shown in Fig.~\ref{fig:Results}d. Fitting the data to an exponential decay model yields a spin-spin relaxation rate of $T_2 =250 (10)$ $\mu$s.

By comparing the magnitude of the echo signal to that of the FID, we can estimate the tip angle $\theta$ achieved in our measurements. For a $\theta$-$\tau$-$2\theta$ pulse sequence, it is straightforward to show that the magnitude of the echo signal (extrapolated to $\tau=0$) is proportional to $\sin^3\theta$, while the magnitude of the FID signal is proportional to $\sin\theta$.  As we increase the external magnetic field $B_0$,   the density of vortices in the superconductor rises , leading to a reduction in the critical current density $j_c$. As a result, the tip angle is expected to decrease. This is confirmed by our measurements, where $\theta$ varies from $53^\circ \pm 6^\circ$ at $B_0=5.6$~T to $16^\circ\pm 2^\circ$ at $B_0=9.95$~T. Using the same pulse lengths $t_A$ used in our experiments, the measured $j_c$ shown in FIG.~\ref{fig:Optimization}e and the calculated $b_{1}$ displayed in FIG.~\ref{fig:OurDevice}c, we estimate the average tip angle as $\langle\theta\rangle = \gamma \langle B_{1} \rangle2t_A$, the factor of 2 resulting from the fact that we are irradiating the central line of an $I=3/2$ nuclear spin.\cite{fukushima_1981}. This gives 37$^\circ$ and 11$^\circ$  at 5.6 T and 9.95 T, respectively. These theoretical values show reasonably good agreement with the experimental results, considering the large uncertainties in some parameters, in particular the distance $d$ between the sample and the coil and the precise rf current applied during our pulses.

To benchmark the sensitivity of our NMR setup against other systems, we use a recently proposed figure of merit: the normalized limit of detection for frequency space (nLOD$_f$) \cite{Lepucki2021}. Indeed, the commonly used SNR is not an ideal metric, as it depends on several factors that are unrelated to the NMR probe itself, such as sample type, number of scans and spectrometer bandwidth. To account for these extrinsic factors, the nLOD$_f$ was introduced and defined as \cite{Lepucki2021}
\begin{equation}
\mathrm{nLOD}_f  = \frac{3 n_s \sqrt{\Delta f}}{\mathrm{SNR}_f\sqrt{N}},
\end{equation}
where $\mathrm{SNR}_f =\mathrm{SNR}/\sqrt{N}$ is the measured signal-to-noise ratio for a single scan in frequency space, and $n_s$ is number of spins in the sample (in mol), $\Delta f$ is the spectrometer bandwidth and $N$ is the number of measurement points per scan. Given the dimension of our sample (150 nm $\times$ 1 mm $\times$ 1.5 mm) and the 80$\%$ natural abundance of $^{11}$B, we estimate that our contains $n_s \sim 38 $ nmol of $^{11}$B spins,  corresponding to $2\times10^{16}$ spins. The SNR is determined as the ratio of the peak NMR signal to $\sqrt{2}\sigma$, where $\sigma$ is the standard deviation of the noise,  yielding $\mathrm{SNR}_f =  0.43$. As a result, we obtain a normalized limit of detection of nLOD$_f = 80$ $\mu$mol Hz$^{1/2}$. 

To compare this value with those of other coils reported in the literature, several factors must be considered. First, to eliminate the dependence on $B_0$, the SNR is typically scaled to 600 MHz by multiplying it by (600 MHz/$f_0$)$^{7/4}$. \cite{Lacey1999} Second, most studies are performed at room temperature on $^{1}$H nuclei, which have a different gyromagnetic ratio ${^1\gamma}$ and spin quantum number $I=1/2$ than $^{11}$B. To account for these differences, we can calculate the effective nLOD$_f$ of our coil for a hypothetical hydrogen sample measured at 600 MHz and 300 K as
\begin{equation}
\mathrm{nLOD}_f^{\mathrm{eff}}  = \mathrm{nLOD}_f\frac{300~\mathrm{K}}{T_s}\frac{{^{11}\gamma}}{{^1\gamma}}\frac{I(I+1)}{0.75}\left(\frac{f_0}{600~\mathrm{ MHz}}\right)^{7/4}.
\end{equation}
This yields nLOD$_f^\mathrm{eff} = 250~\mu$mol Hz$^{1/2}$, providing a standardized measure of our coil’s sensitivity. As noted by Lepucki \textit{et al.} \cite{Lepucki2021}, there is a strong correlation between the $\mathrm{nLOD}_f$ and the linewidth (FWHM) of the measured NMR signal. The most sensitive microcoils are typically characterized by a $\mathrm{nLOD}_f^\mathrm{eff}$ [ $\mu$mol Hz$^{1/2}$] $\lesssim 0.5$ FWHM [ppm]. Based on this criterion, we conclude that despite the unusual and challenging dimensions of a thin, quasi-2D sample, our meander-line coil performs on par with some of the best microcoils designed for 3D samples. Comparing  $\mathrm{nLOD}_f$ with other NMR studies on thin films is challenging, as many of these studies do not report key experimental parameters. 

A simplistic comparison based purely on the number of spins measured shows that we achieve 1000 times more sensitivity than Liu \emph{et al.}~\cite{liu_2017} achieved using a normal-metal surface coil, though the number of scans that they used is not known. Lou \emph{et al.}~\cite{lou_2023}, achieved a measurement of a similar number of spins to this work ($2\times 10^{16}$) at temperatures <40 K using a normal-metal surface coil. However, this feat was accomplished using about 80 times as many scans as our measurement. Despite imperfect tip angles, our measurements at 10 T required only 25,000 scans with a 20 ms repetition time, totaling approximately 500 s of acquisition time while maintaining acceptable SNR. This indicates that our superconducting meander-line surface coil provides a modest advantage over normal-metal surface coils. Furthermore, it allows for the study of samples with smaller lateral dimensions, here $1\times 1.5$ mm$^2$.  A similar number of spins could also be studied in thin-film CrCl$_3$, but this was made possible with ferromagnetic enhancement~\cite{Havemann_2024}. Furthermore, we note that a number of other papers report measurements on a similar number of spins but in a 3d geometry, which requires a different approach from the one adopted here~\cite{lepucki_microcoils_2021,Lepucki2021}.




\section{Discussion}\label{sec:Persp}

The superior sensitivity of our meander-line coil can be attributed to its ability to achieve large inductive coupling with the nuclear spins of very thin samples while maintaining a high quality factor.  This coupling, which is quantified by the filling factor $\eta$ or the factor $\beta_{\mathrm{det.}}$ presented above, arises from the exponential increase in RF field strength near the surface. For our specific coil and sample dimensions, we estimate the filling factor of the meander coil as $\eta_\mathrm{meander} = \int_{V_S} b_{xy}^2 (\vec{r}) d^3\vec{r} /\mu_0 L \sim 0.5\%$. In comparison, an ideal flat solenoid perfectly wound around the sample (1.5 mm in length, 1 mm in width, and 500 µm in thickness, including the silicon substrate) would have a filling factor of approximately $\eta_\mathrm{solenoid} \sim V_s/2V_c = 0.015\%$. Consequently, our meander coil is nearly 50 times more sensitive than this hypothetical solenoid. This result underscores the critical role of inductive coupling in increasing the sensitivity of NMR coils.

In addition to enhancing sensitivity, the large inductive coupling allows for significantly lower RF pulse power compared to measurements with standard RF coils. This, in turn, offers several additional advantages. First, it simplifies the instrumentation,  eliminating  the need for a high-power ($\sim 500$ W) amplifier.  In our experiment, we directly use the output of a standard NMR console. With power to spare, a broadband coupler can replace the conventional setup made of of crossed diodes and a $\lambda/4$ cable, further simplifying the system. This facilitates the incorporation of a cryogenic --- and potentially quantum-limited --- preamplifier to reduce the measurement noise. Additionally, the low power requirement should allow for the use of off-the-shelf electrically-controlled varactors, which would normally be unsuitable for high-power NMR experiments, in place of manually adjustable capacitors. Finally, these advantages could simplify NMR experiments at mK temperatures within a dilution refrigerator, further increasing the system's sensitivity.  

The sensitivity of the current setup is limited in several ways by the critical current density $j_c$ of the superconductor used to fabricate the meander. First, as evidenced by our measurement, the $j_c$ of our NbN film prevents the tip angle $\theta$ from reaching the optimal value of $90^\circ$.  Increasing $j_c$ to enable  $\theta = 90^\circ$ would enhance the SNR by a factor  $\sim 1/\sin{\theta}$, corresponding to a gain of $\sim$ 4 for the lowest measured $\theta$. For spin-echo measurements, the ability to achieve a full $\theta=180^\circ$ spin rotation would result in a dramatic increase in SNR, potentially up to 50 times ($\sim 1/\sin^3\theta$) in spin-echo measurements. Moreover, limited  $j_c$ places strong constraints on the minimum dimensions of the meander. A simple scaling analysis based on the model presented in Sect. \ref{sec:theory} shows that increasing $j_c$ by a factor $\alpha$ while simultaneously reducing the meander period $a$, thickness $t$ and sample-coil distance $d$ by a factor $1/\sqrt{\alpha}$ would maintain the same critical current $\mathcal{I}_c$ but enhance the RF field $b_{xy}$ by a factor of $\sqrt{\alpha}$. This stronger inductive coupling would increase the detection efficiency $\beta_{\mathrm{det.}}$ by the same factor, and also enable tip angles as large as $180^\circ$, leading to a total SNR improvement of $\sim 4\sqrt{\alpha}$ for FID measurements and $\sim 50\sqrt{\alpha}$ for spin-echo measurements, even at very large $B_0$.

A promising route to overcome the $j_c$ limitation and realize these sensitivity gains is by using a high-temperature (high-$T_C$) superconducting film. For instance, 250-nm-thick YBCO films have been shown to sustain $j_c \simeq 30$ MA/cm$^2$ under $B_0 = 5$ T ~\cite{Stangl2021}.This represents a gain of $\alpha \sim 100$ compared to our current NbN-based meander, translating into a projected sensitivity enhancement of $ 4\sqrt{\alpha}=40$. Such an increase would enable the study of significantly thinner films. For instance, using YBCO, it would be possible to measure a boron film of thickness $ 150/\sqrt{\alpha}\approx 4$ nm with comparable SNR to what is reported here. Combined with the other strategies mentioned above, such as lower temperatures and the use of a cryogenic preamplifier, this approach offers a realistic pathway towards achieving measurements in the monolayer limit. Access to a full 180$^\circ$ spin rotation would unlock the full toolkit of NMR pulse sequences, while the higher critical field and critical temperature of a high-$T_C$ superconductors would broaden the accessible experimental parameter space for materials of interest. This would allow us to study the temperature-dependent properties of correlated materials, notably the shift and the spin-lattice relaxation rate, up to temperatures approaching 90 K.  This would open up new opportunities for investigating a wide range of 2D materials and the exotic phases they host, for example enabling an investigation of correlated electronic states in twisted bilayer graphene\cite{cao_2018} and the possible discovery of a Kitaev quantum liquid in exfoliated $\alpha$-RuCl$_3$ and associated heterostructures~\cite{Kitaev2006,Baek2017,yang_2023,massicotte_2024}.
High-$T_c$ meander-line surface coils could also be used to extend the study of Lou~\emph{et al.}~\cite{lou_2023} to study exotic superconductivity and nematicity in even thinner films of FeSe, as well as samples with smaller lateral dimension, while maintaining an acceptable signal-to-noise ratio.

\section{Conclusion}\label{sec:Conclu}

To conclude, we showed that a superconducting meander-line surface coil enables NMR measurements on nanoscale thin films. Despite the early stage of development of this technique, it provides a clear advantage compared to standard RF coils and similar surface coils based on normal metal. We demonstrated this technique on a 150 nm boron thin film and showed that the sensitivity is limited primarily by the critical current density of the superconductor. Based on these results, we proposed a development strategy leveraging high-$T_C$ superconducting meanders to further enhance the technique's sensitivity. This opens a realistic and exciting pathway toward performing NMR measurements on thin films down to the monolayer limit.

The authors would like to acknowledge Stéphane Morin for providing the boron sample, Christian Lupien and Patrick Fournier for helpful discussions and Simon Fortier for technical support. This work was supported by the Canada Research Chair (CRC) program,  the Natural Sciences and Engineering Research Council (NSERC), the Canada First Research Excellence Fund (CFREF), the Fonds de recherche du Québec - Nature et technologies (FRQNT), the Université  de Sherbrooke, and the Canada Foundation for Innovation (CFI).
\begin{acknowledgments}
\end{acknowledgments}




%

\end{document}